# Broadband directional coupling in aluminum nitride nanophotonic circuits

## Matthias Stegmaier[*] and Wolfram H.P. Pernice

*Institute of Nanotechnology (INT), Karlsruhe Institute of Technology (KIT), D-76128 Karlsruhe, Germany*
[*]*matthias-stegmaier@gmx.de*

**Abstract:** Aluminum nitride (AlN)-on-insulator has emerged as a promising platform for the realization of linear and non-linear integrated photonic circuits. In order to efficiently route optical signals on-chip, precise control over the interaction and polarization of evanescently coupled waveguide modes is required. Here we employ nanophotonic AlN waveguides to realize directional couplers with a broad coupling bandwidth and low insertion loss. We achieve uniform splitting of incoming modes, confirmed by high extinction-ratio exceeding 33dB in integrated Mach-Zehnder Interferometers. Optimized three-waveguide couplers furthermore allow for extending the coupling bandwidth over traditional side-coupled devices by almost an order of magnitude, with variable splitting ratio. Our work illustrates the potential of AlN circuits for coupled waveguide optics, DWDM applications and integrated polarization diversity schemes.

## 1. Introduction

The continued attraction for nanophotonics is largely based on the potential for combining waferscale photonic circuits with electronic circuits on a common platform [1]. To date a wide range of such optical circuits have been fabricated in silicon exploiting robust fabrication routines used extensively in electronic circuit manufacturing. In particular after the realization of silicon-on-insulator (SOI) wafers, densely integrated photonic circuits fabricated from silicon thin films have demonstrated excellent performance [2-4]. Due to the relatively small bandgap of silicon of 1.1 eV [5], devices based on SOI are however restricted to operation wavelengths above 1100nm. For wider band applications also covering visible and ultraviolet (UV) wavelengths, photonic devices have to be realized in different material systems with larger bandgaps. Within the group of wide bandgap semiconductors, to date silicon nitride ($Si_3N_4$) has widely been used for the realization of CMOS compatible nanophotonic devices [6-10]. Low material absorption [11,12] and good mechanical properties [13] make $Si_3N_4$ also a prime candidate for opto-mechanical devices. However, the deposition of thick $Si_3N_4$ films of high quality, which are required for nanophotonic waveguides at longer wavelengths, is often accompanied by large internal stress. High quality, stoichiometric silicon nitride films deposited by low pressure chemical vapour deposition (LPCVD) possess internal tensile stress with a magnitude up to 1GPa, which may lead to crack formation when grown to a thickness above a few hundred nanometers [14]. Recently, photonic devices based on GaN [15,16] and AlN [17-19] have been explored as attractive alternatives to conventional CMOS compatible materials. Deposited onto oxidized silicon substrates, these material systems form the analogue of SOI as AlN-on-Insulator (AOI) or GaN-on-Insulator (GOI). Because of a large bandgap of 6.14eV [20] AlN is broadband transparent from 220nm to 13.6μm, thus covering the entire visible wavelength range. Importantly, the large bandgap also enables applications in the ultraviolet wavelength regime, which has triggered renewed interest also with respect to GaN devices of similar function [21]. At the same time, the underlying silicon oxide layer prevents leaking of evanescent waves into the silicon substrate which would lead to absorption below wavelengths of 1100nm. Furthermore, free-carrier effects commonly encountered in silicon photonic circuits are not present, allowing for operating AlN circuits at high input powers. Furthermore, low propagation loss was obtained in AlN ring resonators and non-linear effects such as second harmonic generation and electro-optical modulation have thus been demonstrated [20].

Here we demonstrate essential components for on-chip routing and power splitting on an AlN nanophotonic platform. We implement on-chip beam splitters with variable splitting ratio based on directional couplers. In suitable waveguide geometries, we achieve compact 50/50 splitters with low insertion loss and high repeatability. Combining two directional couplers into an on-chip Mach-Zehnder interferometer, we are able to quantify the coupling uniformity of fabricated devices within the telecoms C-band, which is of importance for DWDM applications and polarization diversity schemes. Over the 3dB-bandwidth of the coupler we obtain splitting uniformity better than 33dB. Since the optimized waveguides support both TE- and TM-polarization, the coupling ratio can be tuned by rotating the polarization of the incoming modes. In order to further extend the coupling bandwidth we design three-waveguide devices that strongly suppress dispersion effects and thus provide coupling uniformity above 20dB over a wavelength range above 60nm. By adjusting the coupling rate between the neighboring extraction waveguides, the splitting ratio can be freely chosen while maintaining high splitting uniformity. Our results allow for robust designs not susceptible to fabrication imperfections and enable broadband evanescent coupling in an integrated framework. Such designs may be used for DWDM applications and also enable novel functionality in optical signal processing and frequency mixing applications.

## 2. Design of dual-waveguide directional couplers

In order to split propagating light into different channels evanescent coupling between neighboring waveguides is commonly exploited. Such co-propagating modes are at the heart of traditional directional couplers employed in modern telecommunication devices. By adjusting the coupling condition between the waveguides, devices with desired properties can be conveniently designed within the framework of coupled-mode theory (CMT).

*2.1 Simulations*

An exemplary fabricated directional coupler embedded in a photonic circuit as described in further detail below is shown in Fig. 1(a). In the middle part of the device two waveguides lie close together as illustrated in Fig. 1(b) leading to mode overlapping and therefore evanescent coupling. The waveguides have a width $w$, height $h$ and are separated by a gap $g$. Furthermore, the sidewall is assumed to be sloped with an angle $\theta$, which is observed during the fabrication of AlN waveguides. Evanescent coupling between the waveguides causes a periodic power exchange characterised by the coupling length $L_c$ which is definded as the distance required for full energy tranfer. Hence, light coupled into the lower input waveguide will be fully transferred into the upper waveguide after $L_c$, leading to zero output optical power in the lower output waveguide. Different splitting ratios can be obtained by selecting a coupling length smaller than $L_c$.

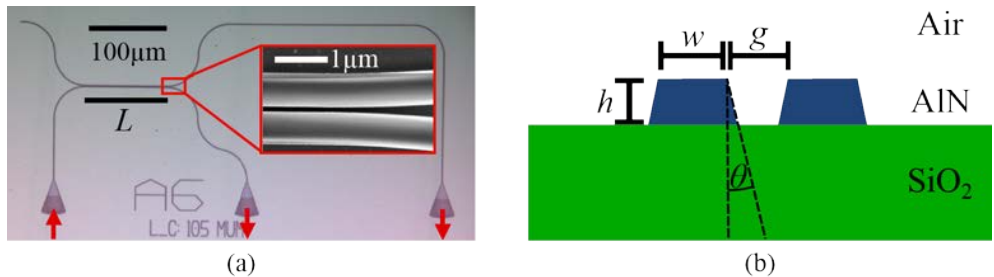

Fig. 1 (a) Exemplary fabricated directional coupler photonic circuit. Light is coupled into and out of the device via grating couplers attached to the input and output ports. In the coupling region in the center of the figure two waveguides are built parallel to each other over a fixed length L. The unused upper left input port is terminated with a taper to scatter incoming light out of the structure and suppress reflections. (b) Cross-sectional view of the coupling area. Two AlN waveguides of width $w$, height $h$ and sidewall angles $\theta$ lie close together, separated by a gap of size $g$, on an insulating $SiO_2$ substrate.

As can be shown by employing CMT [22], $L_c$ depends on the coupling coefficient $\kappa$ but can also be expressed in terms of the effective mode indices of the symmetric $n_{eff,even}$ and antisymmetric $n_{eff,odd}$ supermodes, respectively, where $\lambda$ is the incoming wavelength:

$$L_c = \frac{\pi}{2|\kappa|} = \frac{\lambda}{2(n_{eff,odd} - n_{eff,even})} \qquad (1)$$

The above expression is more convenient than the evaluation of overlap integrals used in CMT, because the effective mode indices can be quickly extracted with a variety of numerical methods. We used this relation to simulate the coupling lengths of our geometry by calculating the different mode indices with the finite-element analysis of COMSOL®. The simulated geometry, presented in Fig. 1(b), consists of two parallel AlN-ridge waveguides of height $h$=500nm, width $w$=850nm, sidewall angles $\theta$=24° and a gap $g$=470nm on top of a SiO$_2$-layer. As refractive index in the telecoms C-band we chose $n_{AlN}$ = 2.12 and $n_{SiO2}$ = 1.44. Because of the relatively thick AlN layer, this waveguide design supports both TE-like and TM-like optical modes, with both symmetric and anti-symmetric mode shape relative to the direction of propagation. For the two polarizations we calculated coupling lengths $L_c^{TM}$=32µm and $L_c^{TE}$=67µm, respectively. The resulting field distributions of the TM-like modes can be seen in Fig. 2(a). The power transfer characteristics are also confirmed with full-vectorial finite-difference time-domain (FDTD) simulations using the commercial package OmniSim, as shown in Fig. 2(b).

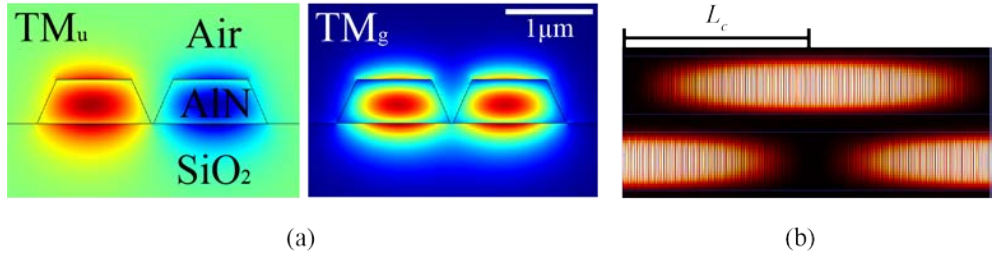

(a) (b)

Fig. 2 (a) Simulated E$_y$-profiles of the symmetric and antisymmetric TM-like supermodes of the studied directional coupler. The calculations are carried out with the finite-element solver COMSOL® (b) Full-vectorial finite-difference time-domain (FDTD) analysis of the power transfer characteristic of the studied directional coupler. The power of the excited lower TM-like mode is coupled into the upper waveguide within a distance $L_c$=32.5µm, the so-called coupling length. With a longer interaction length $L$ than $L_c$, the coupling leads to a sinusoidal oscillation of the guided power between the two waveguides.

In the FDTD simulation complete power transfer is achieved after a coupling length of $L_c^{TM}$=32.5µm and $L_c^{TE}$=61.5µm, consistent with the finite-element simulations. Upon increasing the coupling length, the optical power is transferred back into the input waveguide, leading to a characteristic beating pattern in longer devices.

*2.2 Experimental realization*

In order to verify the simulated results, we fabricate corresponding devices in AOI. Our devices are prepared from an AOI stack, consisting of a highly c-axis oriented polycrystalline AlN film with a thickness of 500nm, which is sputter deposited onto a silicon substrate with a buried oxide layer of 2.6 µm thickness. The preferential crystal orientation is confirmed by the measurement of the full width half maximum (FWHM) value of the rocking curve corresponding to the (002) X-ray diffraction peak, which is found to be below 2 degrees for the material used in our studies. Nanophotonic circuits are defined with electron-beam lithography using a Jeol 5300 50kV system and Fox15 negative tone e-beam resist. Subsequently, the developed samples are etched in Cl$_2$/SiCl$_4$/Ar inductively coupled plasma using an Oxford 100 system. By doing so, we are able to fabricate AOI devices with overall coupling loss below 12dB, corresponding to an overall transmission of up to 6%. The coupling loss consist of 5dB input and output coupling losses and propagation loss of approximately 3.5dB/cm due to residual sidewall roughness, respectively. Even though lower propagation loss has been obtained previously using a Cl$_2$/BCl$_3$/Ar dry etching chemistry [18], the functionality of our devices is not affected by the associated losses.

To measure the performance of the fabricated structures light is coupled into and out of the circuits using focusing grating couplers with a central coupling wavelength in the telecoms C-band [24]. The functionality of the directional couplers at a fixed wavelength is determined by measuring transmission through photonic circuits similar to the one shown in Fig. 1(a). Light from a tunable laser source (New Focus Venturi 6600) is coupled into the lower waveguide using an optical fiber array with three fibers aligned simultaneously to the on-chip grating couplers in the device. After propagation through the circuit, the transmitted signal in the two output ports is recorded with a low-noise photodetector (New Focus, Model 2117). Since the excitation of a specific mode is polarization dependent, we employed a fiber polarization controller to adjust the polarization of the fiber mode which is launched onto the input grating coupler. By doing so, we can selectively excite either of the TE or TM modes and therefore study their different behavior.

We fabricated several directional couplers with varying interaction lengths $L$ in the range between 50µm and 300µm. The obtained coupling efficiencies are plotted versus $L$ for both the TE-like and the TM-like modes in Fig. 3(a),(b). As can be seen, the resulting curves are well described by the sinusoidal oscillation expected for directional couplers. The data are fitted to the theoretically expected curve, yielding coupling lengths of $L_c^{TM}$=33.3+/-0.2µm and $L_c^{TE}$=64+/-1.2µm for the TM-like and the TE-like mode, respectively. These measured values are in good agreement with the simulated coupling lengths of $L_c^{TM}$=32µm and $L_c^{TE}$=67µm.

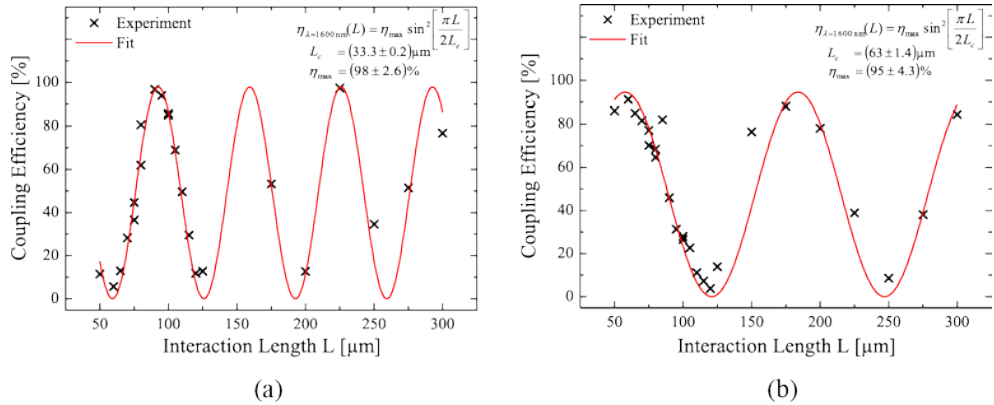

Fig. 3 Coupling efficiencies of (a) the TM-like and (b) the TE-like mode measured with different directional coupler devices with varying interaction length $L$. The obtained data are fitted to the theoretical sinusoidal oscillations yielding coupling lengths $L_c$ consistent with the simulated values of $L_c^{TE}$=32µm and $L_c^{TM}$=67µm. In addition, almost complete power transfer for both modes, i.e. values for $\eta_{max}$ close to 100%, is obtained.

In particular, we note that near complete optical power transfer into both the upper and lower waveguides is obtained for both polarizations. Furthermore, because TE and TM modes yield vastly different coupling lengths, properly designed directional couplers may be employed to separate hybrid TE-TM modes into their respective polarization components.

*2.3 MZI with directional couplers*

To test the applicability of a directional coupler as a beam splitter, a setup more sensitive to the exact power splitting ratio is required than provided by the direct measurement presented in the last section. For that purpose, we designed Mach-Zehnder interferometers (MZI) as shown in Fig. 4(a), which utilize directional couplers as beam splitters. Coming from the left grating coupler, the input power is divided by the first directional coupler into the two arms forming the MZI, which have different length. Upon propagating through the interferometer, the two lightwaves are added with the second directional coupler at the output of the MZI. Then the power transmitted through the second directional coupler depends on both the two incoming mode amplitudes and the phase difference between them. The interference leads to characteristic fringes in the transmission spectrum as shown in Fig. 4(b), with a free spectral range that depends on the path difference of the MZI. Again, we record the transmission through the device in both the upper and lower waveguides, which lead to complementary transmission curves. For symmetry reasons we also include a dummy waveguide on the left directional coupler, which is terminated with a trapezoidal waveguide diffuser in order to prohibit backreflection from the open end (even though only very little light is expected to enter the diffuser end in the first place).

Due to the dispersion of the coupling length and the wavevector inside the AlN waveguides, the ratio of the incoming mode amplitudes and their phase difference is varied when sweeping the incoming wavelength. This way, both constructive and destructive interference take place in the second directional coupler. We measured the corresponding powers, $P_{max}$ and $P_{min}$, and calculated the extinction ratio defined as the logarithmic ratio of these powers for different wavelengths. This quantity is a good measure for the accuracy of the beam splitter since it is sensitive to small deviations from the perfect splitting ratio as shown in Fig. 4(c). As $P_{max}$ and $P_{min}$ occur at different wavelengths and the extinction ratio is defined at one wavelength, the maximal transmitted power $P_{max}$ corresponding to a measured minima $P_{min}$ extracted by interpolation using the transmission envelope as depicted in Fig. 4(b).

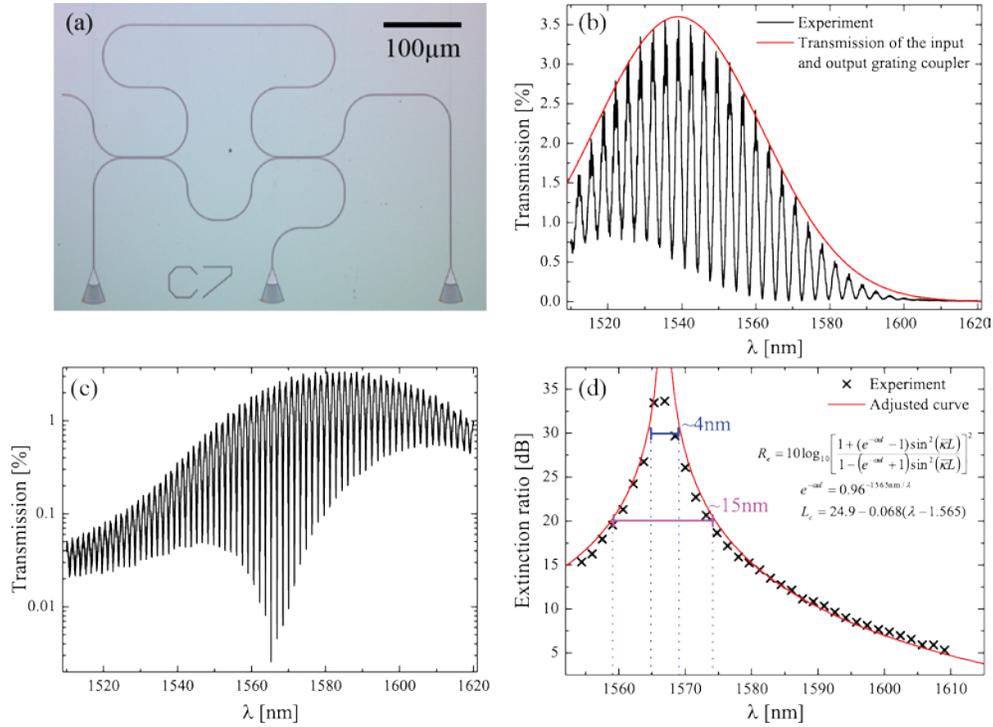

Fig. 4 (a) Optical micrograph of a fabricated MZI design employing directional couplers as beam splitters. The light is launched into the lower left grating coupler, split by the first directional coupler, transferred again by the second directional coupler and finally coupled out via two grating couplers. Fringes corresponding to constructive and destructive interference at the output directional coupler provide a sensitive measure of the quality of the power splitter. (b) Recorded overall transmission at the middle output port. The specially marked Gaussian envelope corresponds to the transmission curve of the input and output grating coupler only. The observed fringes are the result of interference in the reuniting directional coupler (c) Logarithmic plot of another measured transmission curve. The used device had a longer path difference within the MZI than the device used for the upper transmission curve, leading to a smaller fringe separation. (d) Measured extinction ratio at the middle output port, corresponding to the lower waveguide. In the shown exemplary device, a perfect 50:50 splitting ratio is achieved around λ=1567nm. The observed dispersion is a result of the linear decrease of the coupling length $L_c$ with increasing wavelength. Extinction ratio in excess of 30dB is obtained over a 4nm bandwidth.

Results for the device shown in Fig. 4(a) are plotted in Fig. 4(d). The extinction ratio reaches its maximum value of almost 35dB at λ=1567nm and decreases first sharply then slowly with increasing distance from this peak. This sharp decline yields 20dB and 30dB bandwidths of roughly 15nm and 4nm, respectively. The observed curve can be accurately described by the theoretical model using suitably adjusted COMSOL results. For the simulation the waveguide parameter $w=810$nm, $h=500$nm, $g=360$nm and $\theta=11°$ were used, measured from high-resolution SEM images. The resulting dispersion of the coupling length with intercept $L_{c,0} = 24.9$µm and slope $m = -0.068$ is in good agreement with the simulated dispersion with $L_{c,0} = 24.9$µm and $m = -0.078$. While we achieve high extinction ratio on the available wavelengths where destructive interference occurs, even higher values may be possible by designing MZIs with larger path difference and thus a finer sampling rate.

## 3. Beamsplitter model: theory and simulations

The previous section reveals the disadvantage of traditional directional couplers for use as beam splitters: even small dispersion leads to a drastic change in the splitting ratio. Hence, this splitter is only applicable for a narrow spectral band. In addition, due to the sensitivity on the ratio interaction length $L$ to coupling length $L_c$, the desired splitting ratio depends in particular on the accurate choice of $L$. All this is a result of operating the coupler at an inconvenient point of the energy exchange, since the conversion efficiency varies linearly around this point. For example, for a directional coupler operated as a 50:50 splitter the following relation for the conversion efficiency $\eta$ holds:

$$\eta\left(\Delta \frac{L}{L_c}\right) = \eta\left(\frac{\pi}{2} \cdot \left[\frac{1}{2} + \Delta \frac{L}{L_c}\right]\right) \approx \frac{1}{2} + \frac{\pi}{4} \cdot \Delta \frac{L}{L_c} \qquad (2)$$

Here $\Delta L/L_c$ denotes a small deviation of $L/L_c$ from the operating point $L/L_c =1/2$. Hence even small deviations from the right design length result in a deterioration of the device performance. To avoid this problem, an alternative beam splitter design has been proposed [24] based on three coupled waveguides in parallel. The design used by P. Ganguly et al. is presented in Fig. 5 for equal gaps, i.e. using $g=g'$. Due to symmetry, this design always transfers light coming from the left middle input port equally to the outer two ports, independent on the interaction length, the wavelength and the excited mode. Improper choice of the length $L$ then results in insertion loss, rather than deviation from the 50/50 splitting ratio. However, the original design is not capable of splitting light in an arbitrary ratio.

For that purpose, we studied the extended geometry shown in Fig. 5(a), consisting of three equal waveguides in parallel, separated by two freely selectable gaps $g$ and $g'$. Since the outer two waveguides are far apart from each other they can be considered to be uncoupled in first order approximation. In this case, phase-mismatch can be neglected because all three waveguides support the same mode profile and the self-coupling terms are small in comparison with the cross-coupling terms. The latter assumption is necessary and cannot be avoided by redefining the effective mode indices because the middle waveguide is coupled to both the outer ones, leading to a self-coupling term of approximately twice the size of the ones of the outer waveguides. Using these assumptions and labeling the coupling coefficients $\kappa$ and $\kappa'$ CMT leads to the following linear differential equation:

$$\frac{d}{dz}\begin{pmatrix} A \\ B \\ C \end{pmatrix} = -i \begin{bmatrix} 0 & \kappa & 0 \\ \kappa & 0 & \kappa' \\ 0 & \kappa' & 0 \end{bmatrix} \cdot \begin{pmatrix} A \\ B \\ C \end{pmatrix} \qquad (3)$$

Since the coefficients do not depend on $z$, the solution of this equation is the matrix exponential of the above coupling matrix. By diagonalizing this matrix, the solution can be calculated explicitly:

$$\begin{pmatrix} A \\ B \\ C \end{pmatrix}(L) = \frac{1}{1+t^2} \begin{bmatrix} t^2 + \cos(\tilde{\kappa}L) & -i\sqrt{1+t^2}\sin(\tilde{\kappa}L) & t\cdot[\cos(\tilde{\kappa}L)-1] \\ -i\sqrt{1+t^2}\sin(\tilde{\kappa}L) & (1+t^2)\cos(\tilde{\kappa}L) & -it\sqrt{1+t^2}\sin(\tilde{\kappa}L) \\ t\cdot[\cos(\tilde{\kappa}L)-1] & -it\sqrt{1+t^2}\sin(\tilde{\kappa}L) & 1+t^2\cdot\cos(\tilde{\kappa}L) \end{bmatrix} \cdot \begin{pmatrix} A_0 \\ B_0 \\ C_0 \end{pmatrix} \quad (4)$$

Here, $t=\kappa'/\kappa$ denotes the coupling ratio and $\tilde{\kappa}=\kappa'\sqrt{1+t^2}$ is the effective coupling coefficient. If light is coupled into the left middle port of this device (IV: $A_0=C_0=0$), then the power transfer to the outer waveguides leads to a power splitting ratio of:

$$\frac{P_a}{P_c} = \frac{|A(L)|^2}{|C(L)|^2} = \frac{1}{t^2} \quad (5)$$

The above relation illustrates, that the power ratio depends only on the ratio of the coupling coefficients and is thus independent on $L$ and $L_c$. Hence, by adjusting the two gaps appropriately this device can be used as a beam splitter of arbitrary splitting ratio. It should be noted that although the power ratio in the two outer waveguides does not depend on the interaction length, the total transferred power does. However, in contrast to the directional couplers described in the previous section, the device presented here is operated at the point at which the conversion efficiency just varies quadratically with respect to small deviations, which is the point where all the power is transferred. Hence, the functionality does not crucially depend on either the interaction or the coupling length.

While the dispersion of $L_c$ does not affect the splitting ratio, the dispersion of $t$ has to be taken into account. Because $t$ is the ratio of coupling coefficients of two almost equal structures differing only in the gap size, one may expect that $t$ is less affected by dispersion than $L_c$. We confirmed this expectation by evaluating the dispersion curve numerically with COMSOL. We calculated the effective mode indices of a structure like in Fig. 1(b) for different gap sizes and different wavelengths. Using Eq. (1), the corresponding coupling coefficients can be obtained from the different mode indices of the symmetric and antisymmetric supermodes. An exemplary result is shown in Fig. 5(b).

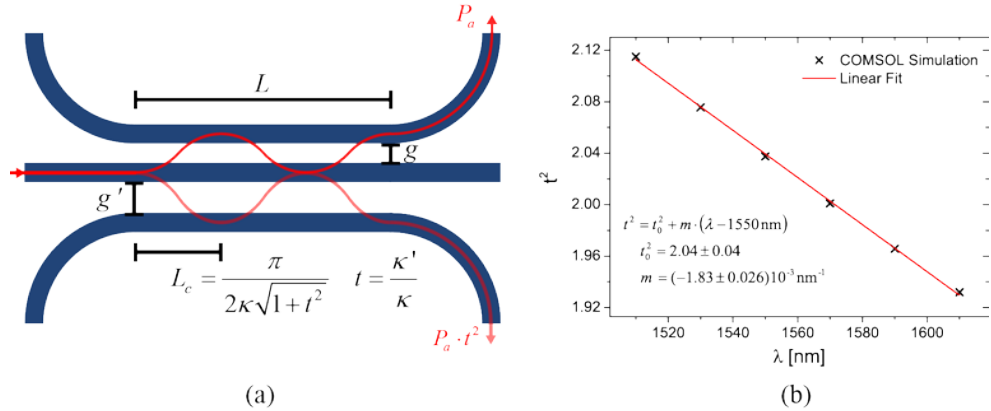

Fig. 5 (a) Sketch of the proposed beam splitter design. The device is composed of three waveguides in parallel with gap sizes $g$ and $g'$. As in common directional couplers, light entering the device at the middle port is transferred to the outer waveguides due to evanescent-wave coupling. The periodicities of this power transfer to the outer waveguides are the same and the ratio of the coupled powers $t^2$ can be adjusted by choosing the gaps appropriately. Note that if the chosen interaction length $L$ is not equal to the coupling length $L_c$ (which is the case in the sketch), some power remains in the middle waveguide. (b) Simulated dispersion of the coupling coefficient ratio $t^2$, carried out with COMSOL®. The wavelength dependence is well described by a linear decrease. Due to a small relative change of $t^2$ across an interval of 100nm, the coupler design is expected to have a broad splitting bandwidth.

As can be seen, $t^2$ changes only slightly from 2.12 to 1.94, corresponding to splitting ratios of 68%:32% to 66%:34%, across a wavelength interval of over 120nm. Therefore the design is only weakly sensitive to fabrication imperfections.

### 4. Three-waveguide beamsplitter: experimental results

In order to accurately quantify the performance of the new splitter design we employ the MZI measurement approach described in the previous section. In the following we present results for two exemplary cases, where the splitting ratio is amenable to the measurement of extinction ratio.

*4.1 Implementation of a 50:50 splitter*

First, we fabricated beam splitters with equal gap sizes as proposed by P. Ganguly et al. To test their 50:50 splitting ratio, again a MZI-design as shown in Fig. 6(a), was used. The incoming light from the lower left input port is transferred to the outer waveguides within the first beam splitter. Power remaining in the middle waveguide is scattered out of the device with a taper. Next, the split light travels through the two arms of a MZI before arriving in the second coupler at the outer waveguides. The power transfer of this device depends on the amplitude ratio and the phase difference between the two input ports. By sweeping the wavelength the phase difference can be altered, leading to constructive or destructive interference. Again the extinction ratio can then be evaluated as a sensitive measure for the uniformity of the splitting ratio.

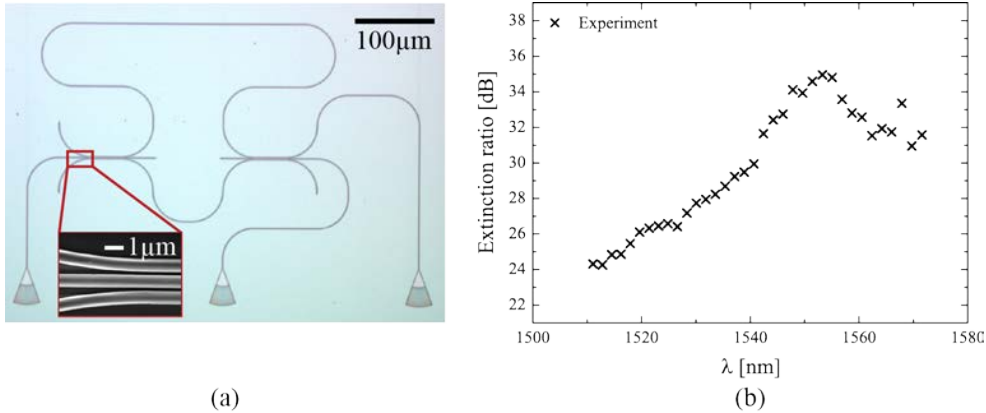

(a)        (b)

Fig. 6 (a) Fabricated MZI based on two 50:50 beam splitters of the proposed design, recorded with an optical microscope. The light enters the device at the left lower grating coupler, is then transferred to the outer waveguides in the first beam splitter, partially reunited in the second one and finally detected after leaving the device via the output grating couplers. (b) Measured extinction ratio of the middle output port, i.e. the middle waveguide of the splitter, versus wavelength of an exemplary device. The reduction of $R_e$ from the expected value of infinity to the observed values of up to 35 dB is mainly a result of non-zero propagation losses. Here, phase mismatch effects caused by different waveguide widths lead to the observed dispersion effect.

The overall transfer function of the studied device can be derived by multiplying the transfer matrices, i.e. the matrix in Eq. (4) for each beam splitter and the matrix of propagation for the middle part. Using the proper initial values ($A_0=C_0=0$), the extinction ratio $R_e$ of the middle output port can then be derived as a function of the absorption coefficient $\alpha$ and the path difference $d$:

$$R_e = 10\log_{10}\left[\frac{1+e^{\alpha d}}{1-e^{\alpha d}}\right]^2 \tag{6}$$

As expected, in absence of absorption the extinction ratio should be infinity independent of the wavelength as a result of the 50:50 splitting ratio being unaffected by dispersion. To estimate the predicted extinction ratio, the path difference $d=500\mu m$ and the approximate propagation loss of 3.5dB/cm in our waveguides are plugged into the formula above, giving a maximal extinction ratio of around 37dB, close to the maximum value measured experimentally.

In Fig. 6(b) the measured extinction ratio of a fabricated MZI is plotted versus wavelength. This particular device is composed of three identical waveguides with width $w=810$nm, height $h=500$nm, sidewall angles $\theta=11°$ and gaps $g=460$nm. In contrast to the theoretical results presented above, $R_e$ shows non-negligible dispersion. The observed residual dispersion is due to a small width difference between the outer the middle waveguide as a result of fabrication errors. This in turn leads to a phase mismatch between the coupled modes and thus reduced maximal conversion efficiency. We note, however, that the results are plotted on a logarithmic scale and thus the deviation from the ideal behavior is still very small.

Though the splitting ratio in the first beam splitter is not affected by any phase mismatch, the maximal power transfer from the outer to the middle waveguide in the reuniting beam splitter does greatly change with it. Hence, the extinction ratio which depends directly on the exact ratio between the maximal and minimal power is changed. Since the phase mismatch depends on wavelength, this phase mismatch most likely causes the observed dispersion. Nevertheless, the measured extinction ratio is higher than 20dB over the whole wavelength interval and even above 30dB over a range of 30nm. Compared with the results obtained with conventional directional couplers, the used beam splitter design gives rise to an 8 times broader bandwidth. We also note that the measured bandwidth of 60nm was limited by the dynamic range of the photodetector combined with the reduced transmission outside the bandwidth of the grating coupler. Hence potentially the coupling uniformity might extend over an even wider wavelength range.

*4.2 Implementation of a 66:33 splitter*

In a second step, we successfully tested the asymmetric design for a power splitting ratio of 66:33, as an example for an arbitrary splitting ratio. This particular splitting ratio is chosen, because the splitter properties can also be quantified using the MZI approach described above. The fabricated design is shown in Fig. 9(a). In the current circuit, the incoming light propagates into the middle port of the beam splitter. Then, it is transferred to the outer two waveguides with a splitting ratio close to 66:33. Before the two arms are reunited, the waveguide carrying the higher power it split using a Y-splitter which results in equal power in both arms. Eventually, the arms are reunited with another Y-splitter. Because the Y-splitter is almost dispersion independent, the obtained transmission results reflect the performance of the directional coupler design. Depending on the power ratio and the phase difference between the two arms, the modes interfere constructively or destructively in the output port (which is the grating coupler in the middle). By sweeping the wavelength, the phase difference can be altered and therefore both cases can be observed.

Using the beam splitter transfer matrix and the appropriate propagation matrices, the overall transfer matrix of the used device can be calculated. From that the output power in dependence of wavelength can be obtained by applying the proper initial values ($A_0=C_0=0$). Eventually, the following expression for the extinction ratio can be derived:

$$R_e = 10\log_{10}\left[\frac{\frac{t}{\sqrt{2}}+e^{\alpha d}}{\frac{t}{\sqrt{2}}-e^{\alpha d}}\right]^2 \qquad (7)$$

Here, $t$ denotes the wavelength-dependent ratio of the coupling coefficients, $\alpha$ is the absorption coefficient and $d$ the path difference. Due to different propagation losses in the two different long arms of the device, the perfect splitting ratio is shifted from 66:33 (corresponding to $t^2=2$) to slightly larger values. If one assumes a path difference $d=500\mu m$ and propagation losses of 3.5dB/cm, the above formula yields the divergence at $t^2=2.17$ corresponding to a power ratio of 68:32.

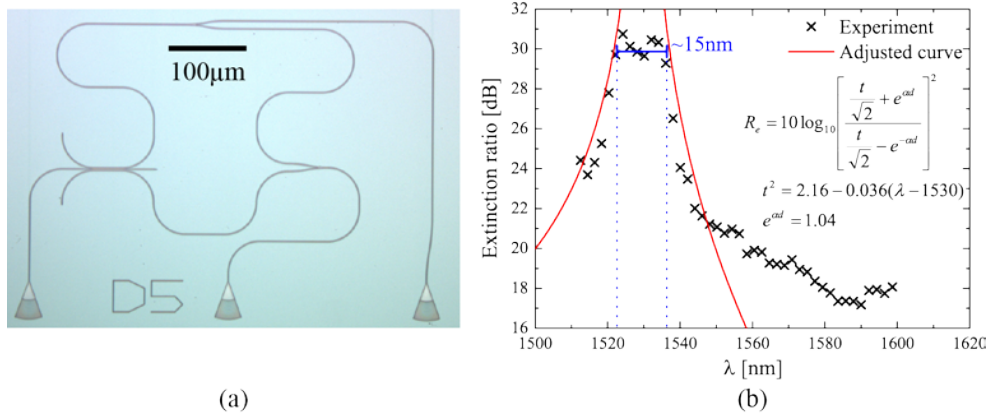

Fig. 9 (a) Optical Micrograph of a fabricated device to test the splitting ratio of the asymmetric 3-waveguide directional coupler. The light coming from the input grating coupler at the lower left is transferred to the outer two waveguides of the beam splitter. Whereas the lower port guides the light directly to the reuniting Y-splitter, the power in the upper waveguide is first split by an additional Y-splitter. If the transferred power ratio is 66:33, then the incident power values in the reuniting Y-splitter are equal due to the bisection of the power in the upper waveguide. Hence, measuring the extinction ratio in the output port is a good measure of the quality of the power splitting. (b) Measured extinction ratio of the middle output port plotted against wavelength for an exemplary device. A perfect 66:33 splitting ratio is achieved around $\lambda=1530nm$ and the obtained 30dB bandwidth is 15nm. Phase mismatch effects and the performance of the used Y-splitters limit the obtained extinction ratios and have an impact on the observed dispersion. Propagation losses cause a shift of the perfect splitting ratio.

The evaluated extinction ratio is plotted in Fig. 9(b) versus wavelength. The presented device has waveguides of width $w=1\mu m$, height $h=500nm$, sidewall angles $\theta=11°$ and gap sizes of $g=470nm$ and $g'=540nm$. The high values of over 30dB close to $\lambda=1530nm$ correspond to an excellent 66:33 splitting ratio. Again, the observed 20dB and 30dB bandwidths of approximately 18nm and 30nm are much broader than the ones obtained with directional couplers, at least by a factor of 2-3. Deviations from the theoretical curve are due to non-uniformities and residual dispersion of the Y-splitter.

## 5. Conclusions

In the current study we have implemented high efficiency power splitters for the emerging AOI platform. In addition to traditional directional couplers consisting of evanescently coupled nanophotonic waveguides, which provide only narrow splitting bandwidth, a novel three-waveguide beam splitter drastically enlarges the obtainable operation bandwidth. In such optimized designs we obtain high extinction ratio above 20dB over a 60nm coupling bandwidth, allowing for both on-chip filtering and polarization selectivity. Furthermore, our approach is not limited to even power splitting, but can also be engineered to desired variable splitting ratio. In particular, with respect to emerging applications in integrated quantum optics, directional couplers represent the chipscale analogue of free-space beam splitters and are thus an essential ingredient for non-classical nanophotonic circuits.

As the AOI platform allows integrated optics in the visible and UV, the presented devices may also be implemented for operation within these regimes. For that purpose, the waveguide widths should be scaled down to prevent the structures to support many modes which may limit the functionality. In addition, the period of the employed grating coupler would have to be reduced to ensure efficient coupling in the desired spectral regime. Though devices with smaller geometrical dimensions may be more challenging to fabricate, both the visible and the UV regime should be within the possibilities of electron beam lithography.

**Acknowledgments**

W.H.P. Pernice acknowledges support by DFG grant PE 1832/1-1. We also acknowledge support by the Deutsche Forschungsgemeinschaft (DFG) and the State of Baden-Württemberg through the DFG-Center for Functional Nanostructures (CFN) within subproject A6.04. The authors further wish to thank Stefan Kühn and Silvia Diewald for assistance in device fabrication.